\documentstyle[12pt,aasms]{article}
\tightenlines

\def\beq{\begin{equation}}
\def\eeq{\end{equation}}
\begin{document}

{\hfill WISC-MILW-94-TH-14}

\vspace{0.5cm}

\title{COMPARING MODELS OF RAPIDLY ROTATING
RELATIVISTIC STARS CONSTRUCTED BY TWO NUMERICAL METHODS}

\author{Nikolaos Stergioulas and John L. Friedman }
\affil{Department of Physics, University of Wisconsin-Milwaukee,
PO Box 413, Milwaukee, WI 53201\\
email: niksterg@alpha2.csd.uwm.edu, friedman@thales.phys.uwm.edu}

\begin{abstract}
We present the first direct comparison of codes based on two different
numerical methods for constructing rapidly rotating relativistic stars. A
code based on the Komatsu-Eriguchi-Hachisu (KEH) method (Komatsu {\it
et al}. 1989), written by Stergioulas, is compared to the
Butterworth-Ipser code (BI), as modified by Friedman, Ipser and
Parker.  We compare models obtained by each method and evaluate the
accuracy and efficiency of the two codes.  The agreement is
surprisingly good, and error bars in the published numbers for maximum
frequencies based on BI are dominated not by the code inaccuracy but by
the number of models used to approximate a continuous sequence of
stars.  The BI code is faster per iteration, and it converges more
rapidly at low density, while KEH converges more rapidly at high
density; KEH also converges in regions where BI does not, allowing one
to compute some models unstable against collapse that are inaccessible to
the BI code.  A relatively large discrepancy recently reported
(Eriguchi et al. 1994) for models based on the Friedman-Pandharipande
equation of state is found to arise from the use of two different
versions of the equation of state.  For three representative equations of
state, the 2-dimensional space of equilibrium configurations is
displayed as a surface in a 3-dimensional space of angular momentum,
mass, and central density. We find, for a given equation of state, that
equilibrium models with maximum values of mass, baryon mass, and angular
momentum are (generically) either all unstable to collapse or are all
stable. In the first case, the {\it stable} model with maximum angular
velocity is also the model with maximum mass, baryon mass, and angular
momentum. In the second case, the stable models with maximum values of
these quantities are all distinct. Our implementation of the KEH method
will be available as a public domain program for interested users.

\end{abstract}


\hspace{0.38cm} PACS: 4.40.Dg, 95.30.Sf, 97.10.Kc, 97.60.Jd

\section{Introduction}

Models of rapidly rotating relativistic stars have been constructed by
a number of different groups, using a few somewhat different numerical
methods (see Friedman (1994) for a bibliography).  Recent work has been
based on the BI code (Friedman et al. 1986 (FIP), 1989; Lattimer et
al.  1990), on the KEH code (Komatsu et al. 1989a, 1989b; Cook et al.
1992, 1994a, 1994b; Eriguchi et al. 1994; Nishida et al. 1992, 1994;
Hashimoto et al. 1994),  and on a code using a finite element method
(Neugebauer \& Herlt 1984; Wu et al. 1991; Neugebauer \& Herold 1992).
No group has used more than one code, and, as we will see,
discrepancies in reported results reflect differences in versions of
the same equation of state (EOS) and in the number of stellar models
used to compute sequences, as well as numerical error intrinsic to the
codes. \footnote{After this paper was written, we received a copy of a
preprint by Salgado et. al. (1994) reporting results from a new code
based on a method different from those listed here.}

The present paper reports a direct comparison of models of
rotating neutron stars constructed with two different codes. Models are
computed using the FIP implementation of the BI method, and corresponding
models are constructed using a code written independently by Stergioulas,
based on the Cook et al. implementation of the KEH method.  We find good
agreement
between corresponding models. Sensitive quantities, such as the radius and
quantities that depend on it, agree to better than 2\%, despite the fact that
the BI code used only 6 spokes (for one quadrant) and 60 radial grid points,
because of limited computer speed at the time it was implemented.
We used three EOSs, G (excessively soft), C, and L (excessively stiff),
to span a range of compressibility larger than that allowed by current
knowledge of neutron-star matter.  Models based on two other EOSs, FP and A,
were also computed to compare them with results given in Eriguchi et al.
(1994).

Our implementation of the KEH method is used to construct the
two-dimensional family of equilibrium models for EOSs C, L and FP.  The
surface of equilibria is ruled by sequences of constant baryon mass and
angular momentum, from which one obtains the boundary of the region
of configurations stable to axisymmetric perturbations.

Our KEH code has been automated to construct various sequences of neutron
star models and will be available as a public domain program.

\section{Outline of the numerical methods.}

The BI and KEH methods are described in detail in Butterworth \& Ipser
(1976) and in Komatsu et al. (1989a) and Cook et al. (1992),
but an outline of their similarities and differences may be helpful (for a
review of stellar structure prior to 1992, see Friedman \& Ipser
1993). Following Bardeen \& Wagoner (1971), both methods use the same
chart, describing the geometry of an axisymmetric rotating star by a
metric of the form $ds^2 = -e^{2\nu} \ dt^2 + e^{2\psi} \bigl( d \phi - \omega
dt\bigr)^2+ e^{2 \mu} \bigl(dr^2 + r^2 d \theta^2 \bigr),$
with Killing vectors $\partial_t$ and $\partial_\phi$ (potentials
independent of $\phi$ and $t$).  Each method solves,
equation-by-equation, the same set of four field equations and the
integrated equation of hydrostatic equilibrium.

Although the gauge and the set of equations used are identical, the
iterative procedure differs in two ways.  To go from the kth to the
(k+1)st approximation to a star, both methods solve one equation at a
time for one potential at a time.   In BI this is done to reduce the
matrix one inverts to a manageable size.  (With present computers,
handling one potential at a time is less  essential, but it may still
be the most efficient method).  KEH is designed for a one-potential
approach and does not seem to allow a variant in which one solves
simultaneously for the set of perturbed potentials.  The BI iteration
differs from a Newton-Raphson iteration only in this truncation.  KEH
observe that one can write the exact second-derivative terms as
flat-space derivative operators acting on redefined potentials. This
allows them to invert the (truncated) linearized operators by using
explicit flat-space Green functions.   One can view their iteration as
departing from Newton-Raphson by omitting the linearized part of the
operator that is not flat.  This relation between the KEH method and
Newton Raphson can be seen as follows:  First, for simplicity, consider
a single nonlinear equation for a single field $\phi$,
\begin{equation}
L\phi =  S(\phi),
\label{exact}\end{equation}
where $L$ is a linear operator, and $S$ is nonlinear.
Given $\phi_k$, one obtains $\phi_{k+1}=\phi_k+\delta\phi$ via Newton-Raphson
by writing
$$L\delta\phi -  S'(\phi_k)\delta\phi = - L\phi_k + S(\phi_k).$$
This is identical to solving
$$L\phi_{k+1} -  S'(\phi_k)\delta\phi =  S(\phi_k).$$
In the KEH method, one solves the equation,
\begin{equation}
L\phi_{k+1}  =  S(\phi_k),
\label{sucsub}\end{equation}
which can be regarded as the result of replacing the linearized
operator $L - S'(\phi_k)$ occurring in the Newton-Raphson scheme by the
truncated version $L$.
\footnote {Our discussion clarifies the relation between KEH and
Newton-Raphson; but the KEH iteration converges to a solution not
because of that relation, but because Eq.~(\ref{sucsub})  is an
example of the method of successive substitution (see, e.g., Nakamura
1993).  That is, when $\phi_k$ given by (\ref{sucsub}) converges, it
converges to an exact solution, because Eq. (\ref{exact}) is exact.}

For each of the three metric potentials that satisfy an elliptic equation,
the KEH iteration conforms to the above description.  In particular,
consider Eq. (4) of KEH, for the potential $f\equiv \rho
e^{\gamma/2}$.  If one were to follow BI, one would obtain the
$(k+1)st$ value of the potential  from the $k$th value of $f$ and from
the set $Q_k$ of other potentials by solving $$ \nabla^2 \delta f -
{\delta S_\rho\over \delta f}  (f_k, Q_k) \delta f = - \nabla^2 f_k +
S_\rho (f_k, Q_k),$$ to find $$f_{k+1} = f_k + \delta f.$$ If one omits
the term ${\delta S_\rho\over \delta f}  (f_k, Q_k)$, the resulting
equation $$ \nabla^2 \delta f = - \nabla^2 f_k + S_\rho (f_k, Q_k),$$
is identical to KEH, $$\nabla^2 f_{k+1} = S_\rho (f_k, Q_k).$$
The other elliptic equations for the potentials are handled by KEH in
essentially the same way.

The second key difference in the methods is in the quantities held
fixed at each iteration: BI fix the polar redshift and angular
velocity, while KEH fix the ratio $r_p/r_e$ of values of the coordinate
$r$ at the pole and equator and the central density.

In our implementation of KEH, the first order derivatives are
approximated by a standard three-point formula. For the second
derivative, however, we found that, if one uses a standard three-point
formula (as in Eriguchi et al. 1994) the metric potential $\mu$
oscillates in most of the interior of the star along a radial spoke.
We overcame the problem by using twice the grid spacing to evaluate
the second derivatives, which resulted in all metric potentials being
smooth functions of radius and angle. If one does not follow the above
procedure to assure the smoothness of the potentials, the errors
introduced in the physical quantities can range from $ \sim 0.5 \%$ in
the mass, and radius to $ \sim 2 \%$ in other quantities, such as the
angular velocity, angular momentum, and redshifts for maximum mass models.

\section{Internal checks and the nonrotating limit of the KEH code}

The accuracy of the physical quantities in a computed model depends on
the grid-size and the domain of integration. BI and Eriguchi et al.
(1994) integrate over a spherical region of radius equal to about twice the
equatorial radius of the star. We integrate over all space using the
coordinate transformation introduced by Cook et al. (1994).
The size of the grid that we use is $129\times 65$ (radial $\times$ angular),
which is similar to the one used by Cook et al. (1994) and Eriguchi
et al. (1994). Doubling the grid-points in both directions did
not change any physical quantity by more than 0.1 \%.

Our code was checked in the nonrotating limit, by constructing
sequences of spherical models and comparing them to models published in
Arnett \& Bowers (1977).  Along a sequence of nonrotating models, the
maximum differences between the two codes in the mass and radius are
$\lesssim 0.3 \%$ and become consistently smaller as one increases the number
of grid points.

\section{Comparison of rotating neutron star models.}

The two codes were first compared for slowly rotating models. A typical
model is shown in Table 1. The agreement in the mass and angular
velocity is $\sim 0.2 \%$ while the metric potentials agree to within
$0.01 \%$ to $0.4 \%$. The two radii differ by $1.4 \%$.
Next, we compare rapidly rotating models, which are near the maximum
mass (reported as the maximum mass models in FIP),
for EOS C, L, and G (Table 1). Here the agreement ranged from
$0.1 \%$ to $0.6\%$ for the masses and from $0.3 \%$ to $1.2 \%$ for
the radii. The angular velocities agreed to $\sim 0.2 \%$ and most
other quantities to $ \lesssim 0.5 \%$, except for some quantities for
EOS G and L which differed by a few per cent. Fig. \ref{omega} compares
results of the BI and KEH codes for
the metric potential $\omega$ as a function of coordinate radius in the
equatorial plane for the two rapidly rotating, EOS C models in Table 1.
The two curves almost coincide, and the agreement in the other three metric
potentials for the above two models was slightly better. FIP estimate the
accuracy of their results to be $ \sim 1 \%$ in the mass, metric coefficients,
density and pressure distributions while the accuracy of the radius and
quantities that depend on it is estimated at $\sim 5 \%$.  The
differences in the above comparisons were somewhat better than this,
implying an accuracy in finding the location of the surface that is
somewhat (but not dramatically) better than the grid spacing. The vast
increase in computer speed since the BI code was written allows a much
finer grid and correspondingly greater accuracy in surface finding.

To compare the convergence for the two codes, we used as input for each
code a fixed initial model and required the code to converge to the
same final model, for which the initial model served as an inaccurate
guess.  The number of iterations required increases considerably for
both codes as one approaches the maximum mass. BI exploits the fact
that stellar structure is dominated by the lowest-order multipoles
$P_l(\cos\theta)$, and uses a grid of size $60\times 6$, with 6 radial
spokes in a quadrant. This means one truncates the multipole expansion
above $l=12$.  Fig. \ref{dgtt} compares the number of iterations needed
to obtain a given accuracy in $ \vert \delta g^{tt} / g^{tt} \vert $
for a rapidly rotating, high density model.  For the same accuracy, KEH
needs a larger grid and a longer time per iteration:  On the DEC Alpha
we used, BI required 0.2 s/iteration, compared to 0.7 s/iteration for
KEH, for grids yielding comparable error.  Because roughly the same
number of iterations were required for the same accuracy for
low-density models, BI was more than three times as fast, but KEH came
into its own for the highest density models, near and above maximum
mass.  KEH requires fewer iterations near the maximum mass model, and
it has additional advantages:  (i) except for the lowest density
models, BI typically requires one to multiply the inhomogeneous terms
by a parameter $c<1$ to obtain a convergent iteration, and time is
spent adjusting $c$ as central density or angular velocity increase
along a sequence (KEH did not require any such parameter in order to
converge, except
for very high density, unstable models constructed with EOS FP).
(ii) Although BI successfully computes the full
range of equilibria for realistic EOSs, KEH is more robust, allowing
one to compute a wider range of models unstable to collapse. (iii) The
code can be easily written to allow the user to specify the grid size.

We compared models computed with the KEH code to corresponding models
in Cook et al. (1994) and in Eriguchi et al. (1994) and found them in
good agreement (maximum 1-2 \% differences in physical quantities).
\footnote{This agreement is equivalent to that just reported by Salgado
et al. (1994) with whom our KEH code also agrees to within 1-2\%.}
Eriguchi et al. (1994) compared their results to Friedman et al. (1989)
and found similarly good agreement except for EOS FP (Friedman \&
Pandharipande 1981).  For this EOS, the two $\Omega_K$ vs. $M$ curves
substantially disagreed (where $\Omega_K$ is the angular velocity at the
mass-shed limit); there was a 5 \% discrepancy in the
maximum angular velocity and a larger (25 \% ) difference in the central
densities of the maximum-mass configurations.  This seemed worrisome,
but the difference between the computed models turns out to be almost
entirely due to a difference in the version of the EOS used.  Using our
KEH code, we recomputed the $\Omega_K$ vs. $M$ curve for EOS FP, and
found it to be close to the corresponding curve in Friedman et al.
(1989)), (Fig. \ref{eriguchi}).  A computation of the same curve using the
version communicated to us by Eriguchi of the FP EOS gave a curve that
is similarly close to that of Eriguchi et al (1994).  For the version
of FP used in FIP, models with the same central density based on the
KEH and BI codes agree in all quantities to within 3\% (which includes the
error in determining $\Omega_K$).  There is a larger difference in
the value of $\epsilon_{\rm c}$ at the maximum mass model, because the error
in $\Omega_K$ is multiplied by a large value of $\partial
\epsilon_c/\partial \Omega_K$ (where the derivative is evaluated along a
sequence of models at the mass-shed limit).

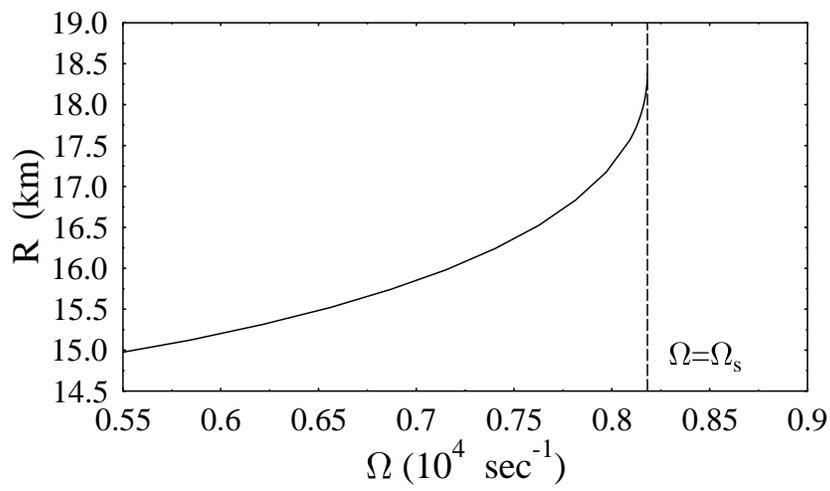
\begin{figure}
\caption{The radius R of a star increases sharply as the star approaches the
mass-shed limit. The sequence of models is that of Fig. 4.}
\label{R_Omega}
\end{figure}

The frequency $\Omega_{\rm s}$ of a satellite in circular orbit at the
star's equator decreases monotonically as $\Omega$ (and hence the
radius of the star) increases, with
$|d\Omega_{\rm s}/d \Omega|_{\epsilon_{\rm c}}$ and
$|d\Omega_{\rm s}/d \Omega|_{M_0}$
apparently diverging at $\Omega=\Omega_{\rm s}\equiv\Omega_K$ (Fig.
\ref{Os_Omega}). One can
exploit the steep slope to accurately determine $\Omega_K$.  Because
the apparent divergence in $|d\Omega_{\rm s}/d \Omega|$ arises from a
similar divergence in $|dR/d \Omega|$ (Fig. \ref{R_Omega}) one needs
that high accuracy to determine $R$ and quantities which depend on it.

\section{Surfaces of equilibria}

   The equilibrium models for a given EOS comprise a 2-dimensional
surface.  Our KEH code is automated to construct sequences of constant
angular momentum and sequences of constant baryon mass, terminating at
$\Omega=\Omega_K$. For a given EOS, we first construct the nonrotating
sequence and the $\Omega=\Omega_K$ sequence. The code then uses this
information to construct a specified number of constant $J$ and
constant $M_0$ (baryon mass) sequences, ruling the 2-dimensional
surface of equilibria.

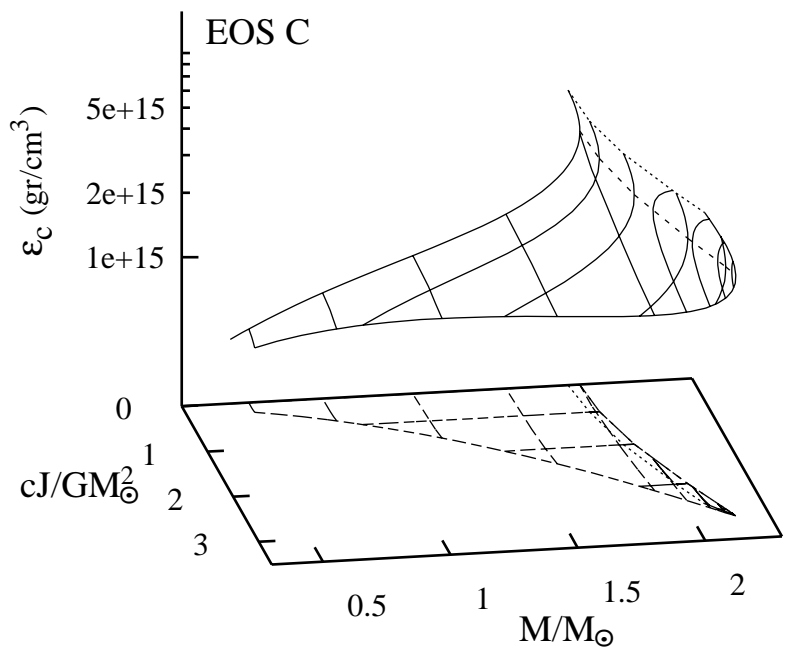
\begin{figure}
\caption{The 2-D surface of equilibrium models is embedded in the
$J$-$M$-$\epsilon_c$ space. The surface is bounded by the spherical
($J=0$) and Keplerian ($\Omega=\Omega_K$) limits and formed by constant
$J$ and constant $M_0$ sequences (solid lines). Also shown are the
axisymmetric instability sequence (short-dashed line), the projections
on the $J$-$M$ plane (long-dashed lines), and the overlapping of the
surface in the $J$-$M$ plane (dotted lines).}
\label{emj_c}
\end{figure}

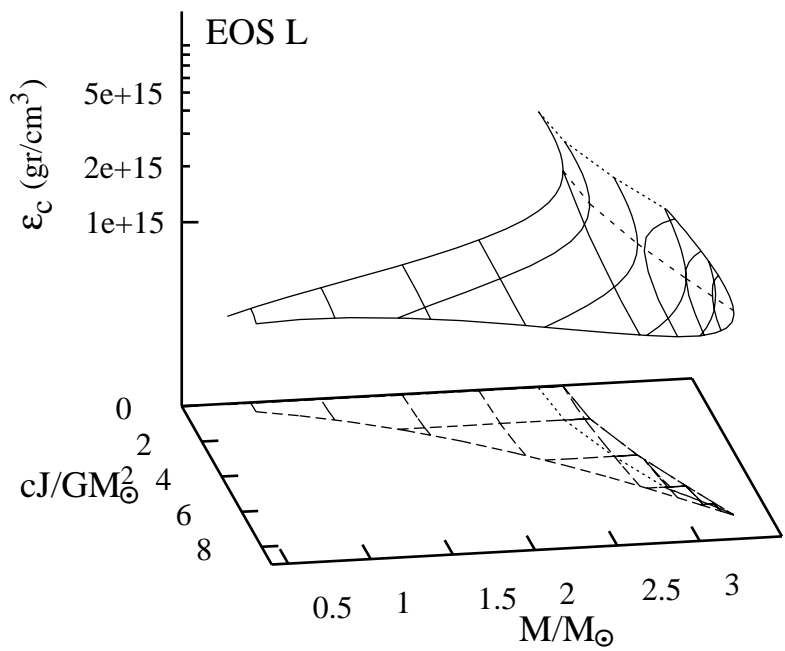
\begin{figure}
\caption{Surface of equilibrium models for EOS L.}
\label{emj_l}
\end{figure}

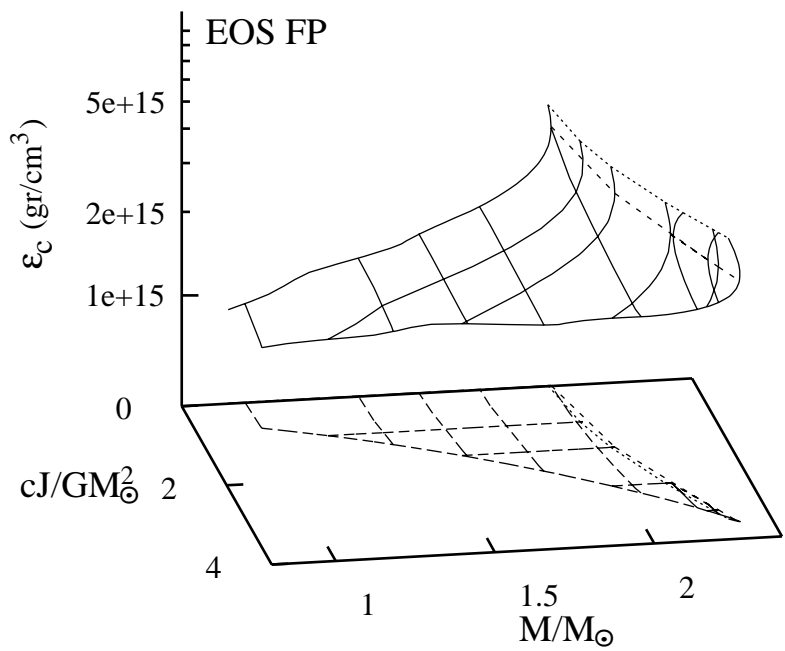
\begin{figure}
\caption{Surface of equilibrium models for EOS FP.}
\label{emj_fp}
\end{figure}

The corresponding surfaces for EOSs C,L and FP are shown in Figs.
\ref{emj_c} to \ref{emj_fp}.  These equations of state are
substantially different, leading to stars whose maximum values of $J$
differ by more than a factor of two.  But their ruled surfaces of
equilibria are strikingly similar, and the similarity seen in the
figures persists for the other equations of state we have examined.  On
each surface, the ridge of maximum mass at fixed $J$ (or minimum mass
at fixed baryon mass $M_0$) marks the onset of axisymmetric instability
(instability to collapse) (Friedman, Ipser \& Sorkin 1988, Cook et al.
1992).  Stars stable against collapse lie on the low-density side of
the dashed line on each surface.  The surfaces fold over at the ridge,
and the diagrams display a sharp contrast between an apparent
cusp at the maximum mass configuration in the $J-M$ plane and the
smooth unprojected boundary of the surface itself.

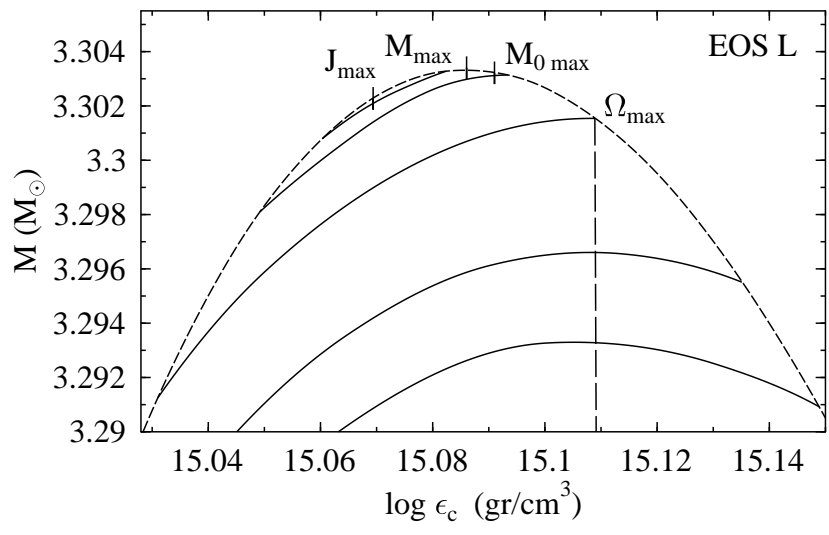
\begin{figure}
\caption{Detail near $M_{\rm max}$ of the mass vs. central density plot for
EOS L. Four different stable models are associated with maximum values of
 $M$, $M_0$, $J$, and $\Omega$. Solid lines are constant $J$ sequences,
the short-dashed line is the mass-shed limit, and the long-dashed line is
the axisymmetric instability sequence. A 257 $\times$ 129 grid was used
for high accuracy.}
\label{jmax}
\end{figure}

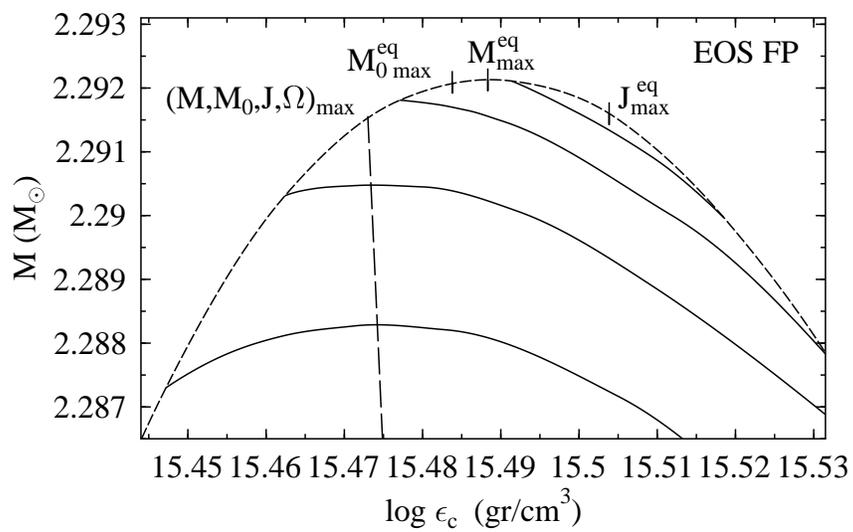
\begin{figure}
\caption{Same as Fig. 9, but for EOS FP. A single stable model has maximum
values of $M$, $M_0$, $J$, and $\Omega$.}
\label{jmax_fp}
\end{figure}

   FIP noted that among all stable models, the model with maximum mass
appeared also to have the maximum $\Omega$, $J$, and $M_0$, and this
apparent coincidence can be seen in Figs. \ref{emj_c} to \ref{emj_fp}.
As Cook et al.  (1992,1994a,b) have found, there is often a slight
difference between the models of maximum $M$ and $\Omega$, and a check
of their claim for EOSs L and FP is shown in Figs. \ref{jmax} and
\ref{jmax_fp}, which depict the region very near the maximum mass
model. In addition, we found that models with maximum values of $J$ and
$M_0$ can also fail to coincide exactly with the maximum $\Omega$ model
and this can be seen from the following argument: The models with
maximum $\Omega$ and maximum $M$ among all equilibria do not coincide.
Thus either $\epsilon_c(\Omega_{\rm max}) < \epsilon_c(M_{\rm max})$ or
$\epsilon_c(\Omega_{\rm max}) > \epsilon_c(M_{\rm max})$. In the first
case, stable models with maximum $M$, $\Omega$, $J$ and $M_0$
coincide. In the second case, $ ( \partial M /\partial \epsilon_c)_J >
0$ for any  $\epsilon_c < \epsilon_c(\Omega_{\rm max})$, and $J(M_{\rm
max}) > J(\Omega_{\rm max})$. Furthermore, $\partial M / \partial
\epsilon_c =0$ at the maximum mass model. Hence, there exists
$\epsilon_c < \epsilon_c(M_{\rm max})$ with $J > J(M_{\rm max})$.
A similar argument, using $( \partial M /\partial \epsilon_c)_{M_0} $,
yields $\epsilon_c(M_{\rm 0 max}) \neq \epsilon_c(M_{\rm max})$, (see Figs.
\ref{jmax} and \ref{jmax_fp}).

Depending on the central density of the fastest rotating stable model
we can distinguish two classes of equations of state: if the fastest rotating
stable model is at a lower density than the maximum mass equilibrium model,
the {\it equilibrium} models with maximum $M$,$M_0$, and $J$ are all unstable
(e.g. EOS FP, Fig. \ref{jmax_fp}); otherwise they are all stable (e.g. EOS L,
Fig. \ref{jmax}). We note that in the first case, the fastest rotating
{\it stable} model is also the configuration with maximum $M$,$M_0$, and $J$,
while in both cases the four different {\it equilibrium} models are all
distinct.
Even when the coincidence of stable models is not precise, the models with
maximum $M$, $\Omega$, $J$ and $M_0$ are generally very close in central
density
(with the exception of EOS M (see Cook et al. 1994b)).
As a result, they {\it appear} to coincide with the sharp edge in the $M$-$J$
plots (Figs.  \ref{emj_c} to \ref{emj_l}).

\vspace{1cm}

We thank Y. Eriguchi for useful discussions and for providing us with his
version of EOS FP and detailed numerical results for selected models. This
research has been supported by NSF grant PHY-9105935.

\clearpage

\clearpage

\begin{figure}
\caption{Comparison of the metric potential $\omega$ at the equatorial
plane for the two rapidly rotating, EOS C models shown in Table 1. The
horizontal axis shows the coordinate radius $r$, while the values of
$\Omega$ and $r_e$ of the model constructed with the KEH code were used
to scale the two axes. The two curves almost coincide.}
\label{omega}
\end{figure}

\begin{figure}
\caption{The fractional change in $g^{tt}$ is plotted against the
number of iterations for a high density, EOS C model. The solid line
was obtained with the BI code and the dashed line with the KEH code.}
\label{dgtt}
\end{figure}

\begin{figure}
\caption{Angular velocity vs. mass plot depicting sequences of models at the
mass-shed limit for EOSs C, A, and FP. Solid lines are present results, long
dashed lines are Eriguchi et al. (1994) results, and short dashed lines are
FIP results. FP (E) is the version of the FP EOS that Eriguchi et al. used.}
\label{eriguchi}
\end{figure}

\begin{figure}
\caption{The frequency $\Omega_s$ of a satellite in circular orbit at
the equator drops sharply as the star approaches the mass-shed limit.
The above sequence of models was constructed with EOS L and with
constant $\epsilon_c=1.21 \times 10^{15} {\rm gr/cm^3}$.}
\label{Os_Omega}
\end{figure}

\end{document}